%% file: main.tex
\documentclass[sigconf, nonacm]{acmart}

\usepackage{balance} 
\usepackage{xspace}
\usepackage{subcaption}

\title[]{The Cost of Classical Multi-Agent Path Finding}

\author{Alvin Combrink}
\affiliation{
    \institution{Chalmers University of Technology}
    \city{Gothenburg}
    \country{Sweden}}
    \email{combrink@chalmers.se}

\author{Sabino Francesco Roselli}
\affiliation{
    \institution{Chalmers University of Technology}
    \city{Gothenburg}
    \country{Sweden}}
    \email{rsabino@chalmers.se}

\author{Martin Fabian}
\affiliation{
    \institution{Chalmers University of Technology}
    \city{Gothenburg}
    \country{Sweden}}
    \email{fabian@chalmers.se}

\begin{abstract}
\input{sections/abstract}
\end{abstract}

\keywords{Classical MAPF, Continuous-time MAPF, Optimality}

\newcommand{\BibTeX}{\rm B\kern-.05em{\sc i\kern-.025em b}\kern-.08em\TeX}

\input{newcommands}

\begin{document}


\maketitle 


\section{Introduction}
\label{sec:Introduction}
\input{sections/introduction}

\section{Background}
\label{sec:Background}
\input{sections/background}

\section{The Cost of the Classical MAPF Assumptions}
\label{sec:ClassicalAssumptions}
\input{sections/classical_assumptions}

\section{Study Design}
\label{sec:StudyDesign}
\input{sections/study_design}

\section{Results}
\label{sec:Results}
\input{sections/results}

\section{Discussion}
\label{sec:Discussion}

\input{sections/discussion}

\section{Conclusion}
\label{sec:Conclusion}
\input{sections/conclusion}




\begin{acks}
Computational resources on the Arrhenius supercomputer were provided by the National Academic Infrastructure for Supercomputing in Sweden (NAISS), funded by the Swedish Research Council.
\end{acks}



\balance
\bibliographystyle{ACM-Reference-Format} 
\bibliography{Bib}


\end{document}

%% file: sections/abstract.tex
Multi-Agent Path Finding (MAPF) is the problem of planning conflict-free paths for multiple agents in a shared space, each from its start to its goal.
Classical MAPF has been the dominant formulation for many years, 
with its assumptions of discrete time and graph-based conflicts presumably easing the search for solutions.
These assumptions limit the physical environments and agents for which a solution is truly collision-free,
and also place an upper bound on solution quality that no algorithmic improvements can lift.
This work investigates how much solution quality, and in what contexts, the classical MAPF formulation forfeits.
Continuous-time MAPF (\MAPFR) relaxes these assumptions, making it a natural counter-formulation to compare against
across various agent counts and sizes, and graph connectedness, topologies, and resolutions.
We find
that continuous time and agent shape consideration are worth relatively little on their own;
their value comes from enabling an expanded range of move actions, 
on average improving solution quality by at least $5\%$ on narrow and constrained maps and $17\%$ on maps with open spaces.
In some cases, the improvements exceed $20\%$.
Doubling the map resolution with classical MAPF recovers less than $3\%$, meaning that little of what is forfeited can be bought back through more compute. 
This work therefore provides insight on when classical MAPF is a reasonable simplification, 
and when \MAPFR unlocks significantly higher-quality solutions.

%% file: newcommands.tex
\newcommand{\repositoryURL}{\url{https://github.com/Adcombrink/classical-vs-continuous-mapf}\xspace}

\newcommand{\MAPFR}{MAPF$_R$\xspace}

\newcommand{\OCCBS}{OC-CBS\xspace}
\newcommand{\AOCCBS}{AOC-CBS\xspace}

\newcommand{\Real}{\mathbb{R}}
\newcommand{\tuple}[1]{\langle#1\rangle}

\newcommand{\Graph}{\mathcal{G}}
\newcommand{\Vertices}{\mathcal{V}}
\newcommand{\Edges}{\mathcal{E}}
\newcommand{\Agents}{\mathcal{A}}

\newcommand{\agent}{a}
\newcommand{\vertex}{v}
\newcommand{\edge}{e}
\newcommand{\startvertex}{s}
\newcommand{\goalvertex}{g}

\newcommand{\DTopt}{\ensuremath{\mathrm{DT}_\mathrm{opt}}\xspace}
\newcommand{\CTLB}{\ensuremath{\mathrm{CT}_\mathrm{LB}}\xspace}
\newcommand{\CTUB}{\ensuremath{\mathrm{CT}_\mathrm{UB}}\xspace}
\newcommand{\Red}{\ensuremath{\Delta}\xspace}
\newcommand{\RedLB}{\ensuremath{\Red_\mathrm{LB}}\xspace}
\newcommand{\RedUB}{\ensuremath{\Red_\mathrm{UB}}\xspace}

\newcommand{\mapEmpty}{\emph{Empty-32-32}\xspace}
\newcommand{\mapEmptySmall}{\emph{Empty-16-16}\xspace}
\newcommand{\mapRoom}{\emph{Room-64-64-8}\xspace}
\newcommand{\mapWarehouse}{\emph{Warehouse-10-20-10-2-2}\xspace}
\newcommand{\mapMaze}{\emph{Maze-32-32-2}\xspace}

%% file: sections/introduction.tex
\emph{Multi-Agent Path Finding} (MAPF) is the problem of planning agent movements on a graph, to get them from where they are to where they need to be without conflicting with each other.
This is often paired with an objective function defining a solution's quality by way of a \emph{cost}, with optimal solutions minimizing the cost.
The applications are many, 
including warehouses and factories~\cite{wurman2008coordinating, MAPF_buzzword, brown2020optimal, varambally2022mapf},
video games~\cite{snape2012reciprocal, Ma_Yang_Cohen_Kumar_Koenig_2017, harabor2022benchmarks}, 
road traffic~\cite{LiAAAI23, goenawan2025astmautonomoussmarttraffic}, and airport management~\cite{morris2016planning, li2019departure}.
The overwhelming majority of MAPF research adopts the \emph{classical MAPF} formulation~\cite{SurveyStern},
for which decades of sustained research has enabled algorithms to find 
optimal solutions for the low hundreds of agents~\cite{CBS,ICBS,li2019disjoint, li2019symmetry}, 
and suboptimal~\cite{ma2019searching, zhang2022learning}, bounded-suboptimal~\cite{li2021eecbs, barer2014suboptimal}, and anytime algorithms~\cite{li2022mapf, LaCAM, okumura2024lacam3} to the thousands.
Scalability is bounded in principle by NP-hardness~\cite{NPhardness, NPHardYakovlev}, yet the proliferation of increasingly scalable algorithms suggests we have not yet reached these limits.
Solution quality, on the other hand, is bounded absolutely by the chosen problem formulation;
no amount of algorithmic progress can produce a higher-quality solution while remaining in the formulation's solution space.
The dominance of classical MAPF is historically well-founded since the assumptions it makes 
supposedly ease the search, and for much of the MAPF field's history, more faithful formulations were simply intractable at useful scales. 
Simply put, a solution to a simplified model was better than no solution at all.

In the classical formulation, 
time is discrete, 
graphs are limited to unit-length edges,
and agents conflict when simultaneously occupying the same vertex or traversing the same edge at the same time~\cite{SurveyStern}.
If agents model real robots in a shared physical space, 
this formulation restricts how and to what resolution the environment can be modelled, 
leaves agent shapes implicitly bounded in size to the graph, 
and incurs artificial delays that consequently inflate the solution cost (we elaborate on this in Section~\ref{sec:ClassicalAssumptions}).
\emph{Continuous-time MAPF} (\MAPFR)~\cite{CCBS} is perhaps the most similar formulation to drop these assumptions, 
such that agents can traverse edges at any real-valued time,
these traversals or waits at vertices taking any positive real-valued duration,
and conflicts considering the agents' shapes.
\MAPFR can therefore model a wider range of physical systems, including agent movements for which classical solutions cannot guarantee collision-freeness.
\MAPFR does not lift all limiting assumptions, such as agents only moving along a discrete set of trajectories instead of in the entire continuous space; 
we do not investigate the cost of those assumptions here.

Classical MAPF nevertheless remains the wide-spread formulation.
\MAPFR on the other hand has received relatively little attention. 
However, recent advances~\cite{PSIPP, CPLP, CCBS, ImprovingCCBS, OCCBS, AOCCBS} pushing \MAPFR to hundreds of agents demonstrate that \MAPFR scalability also responds to research effort.
This makes the following question increasingly relevant:
\begin{quote}
    \textbf{What solution quality does the classical MAPF formulation forfeit relative to \MAPFR, and in what contexts is this forfeit substantial?}
\end{quote}
Existing work only partially answers this question.
Using the \MAPFR algorithm CCBS~\cite{CCBS} and variants of ICTS~\cite{ijcai2018p74},
solution quality improves by $5$--$25\%$ on average when adding non-cardinal edges (such as diagonals) to the map.
Any-Angle MAPF~\cite{AAMAPF} maximises these non-cardinal edges, 
for which Any-Angle CCBS~\cite{AACCBS} provides an illustrative example of such a solution having ${\sim}22\%$ lower cost compared to only using cardinal edges.
CE-CBS~\cite{CE-CBS}, a discrete-time inexact method addressing the related problem of continuous \emph{space} rather than continuous \emph{time}, reports solution quality increasing as agent radius decreases.
Finally,~\cite{GRACE} compares the exact classical MAPF algorithm CBS~\cite{CBS} with several inexact \MAPFR methods, finding lower-cost \MAPFR solutions on only some problems.

These existing works do not fully address the question:
ICTS~\cite{ijcai2018p74} relies on time discretization and reports results for only a set number of agents;
CCBS is a suboptimal~\cite{CCBSrevisit, CBSAA} (although often near-optimal~\cite{OCCBS}) algorithm and reports only aggregated results;
and Any-Angle CCBS's correctness guarantees build on CCBS's without, in our opinion, delivering sufficient theoretical proof that it does not inherit CCBS's inexactness.
A suboptimal \MAPFR algorithm improving on the optimal classical MAPF solution still does not reveal the true magnitude of the gap, 
and since a classical MAPF solution is a valid \MAPFR solution\footnote{Assuming both problems use the same map and that the classical solution respects the requirements for conflict-freeness in the real system (see Section~\ref{sec:ClassicalAssumptions}).}, 
not improving on the optimal classical solution says only something about the algorithm that was used.  
Therefore, these existing studies work only with suboptimal \MAPFR solutions, adjacent problems, 
or without direct and comprehensive comparisons with the classical formulation.

Recent work provides an exact \MAPFR algorithm, \OCCBS~\cite{OCCBS}.
Our preliminary comparisons between \OCCBS and CBS, however, revealed \OCCBS's relatively poor scalability to impede a comparison 
at sufficiently high agent counts within a reasonable runtime. 
\AOCCBS~\cite{AOCCBS} is an anytime-optimal \MAPFR algorithm, building on \OCCBS, which early on provides an incumbent solution and progressively refines it until eventually converging on an optimal solution.
Although \AOCCBS is comparable to \OCCBS for finding optimal solutions, 
it importantly provides bounds on the optimal cost,
thereby enabling a comparison between the classical and \MAPFR formulations at higher agent counts without necessarily converging.

This work contributes the following. 
First, we give a direct comparison between the optimal cost of classical MAPF and \MAPFR. 
Using \AOCCBS's bounds rather than requiring an optimal \MAPFR solution allows the comparison to reach higher agent counts while still keeping the comparison on the formulations rather than on the algorithms.
Second, we quantify this cost difference over discrete versus continuous time, agent size, and map connectedness, on four maps ranging from open spaces to narrow corridors.
We also test whether refining the map resolution allows classical MAPF to recover the loss.
Third, we identify the contexts in which classical MAPF remains an appropriate simplification, and those in which the forfeited solution quality motivates further \MAPFR scalability efforts. 
Our repository\footnote{\repositoryURL} provides the full results and settings.

%% file: sections/background.tex
We formally introduce the classical MAPF and \MAPFR formulations, as well as their respective algorithms CBS and \AOCCBS used here.

\subsection{Classical MAPF}

The classical MAPF formulation~\cite{SurveyStern} is posed on an undirected graph $\Graph = \tuple{\Vertices, \Edges}$ with vertices $\Vertices$ and edges $\Edges\subseteq\Vertices\times\Vertices$.
Each agent $\agent\in\Agents$ is initially at a unique start $\startvertex_\agent\in\Vertices$ and must reach a unique goal $\goalvertex_\agent\in\Vertices$.
Time is discrete; 
at every timestep, each agent either waits at its current vertex for another timestep (wait action), or moves along an adjacent edge to arrive at a neighbouring vertex at the next timestep (move action). 
Two agents conflict if they are located at the same vertex or traverse the same edge at the same timestep.
A solution consists of one plan for each agent $\agent\in\Agents$, with $\agent$'s plan specifying a sequence of actions that start at $\startvertex_\agent$, end at $\goalvertex_\agent$, and do not conflict with any other plans.
There exist many ways to rank solutions. 
Here, we use the \emph{Sum-of-Costs} (SOC):
the cost of a plan is its duration (timestep of the last move action); 
SOC is the sum over all plan costs, which an optimal solution minimises.

\subsection{Continuous-time MAPF}

The \MAPFR formulation is originally introduced in~\cite{CCBS} and further generalised in~\cite{AOCCBS}.
For brevity, we present a specified version in 2D space with homogeneous and circular agents.

Compared to the classical formulation, time in \MAPFR is no longer discrete:
move actions can happen at any real-valued time and take any positive time to complete, and agents can wait at vertices for any positive duration.
Each vertex $\vertex\in\Vertices$ maps to a specific position in $\Real^2$ where an agent waiting at $\vertex$ is located.
Similarly, an edge $\edge\in\Edges$ defines a trajectory $\Real\mapsto\Real^2$ mapping from time to a position where an agent traversing $\edge$ is located. 
Conflicts now consider the agents' shapes, with two agents conflicting if their shapes overlap.
Furthermore, since conflicts are defined by the agents' positions and shapes instead of their location on the graph, changing the shapes also changes what an optimal solution is. 
For instance, multiple agents can traverse the same edge simultaneously, separated by just enough space to not conflict.

\subsection{Conflict-Based Search}
CBS~\cite{CBS} is a classical MAPF algorithm that searches on two levels to find an optimal solution.
A high-level search is done in a binary constraint tree (CT): 
every CT node represents a candidate solution and a set of constraints; 
every plan in a node's candidate solution is found using a low-level planner and has minimum duration while satisfying the node's constraint set.
The CT initially contains only the root with an empty constraint set, with every agent plan representing the shortest path without considering conflicts with other agents.
At every iteration of the search, the CT leaf node with the minimum candidate solution cost is selected for expansion.
If this node's candidate solution is a solution, then it is returned as the optimal.
Otherwise, the node is expanded by 
spawning two child nodes that inherit the current node's constraint set, 
selecting a conflict,
creating one constraint for each agent involved in the conflict,
and adding each constraint to a respective child node.
The candidate solutions at these new child nodes are replanned by the low-level planner to satisfy the expanded constraint set.
This method is guaranteed to produce an optimal solution if one exists.

\subsection{Anytime-Optimal Continuous-time CBS}

\AOCCBS~\cite{AOCCBS} is an anytime-optimal \MAPFR algorithm that early on produces an incumbent solution, and over the algorithm's runtime improves the solution to eventually return the optimal, all while providing bounds on the optimal cost.
It is based on \OCCBS which in turn is based on the continuous-time version of CBS, CCBS~\cite{CCBS}. 
\AOCCBS's search is therefore similar in structure to CBS, consisting of a high-level search in the CT and a low-level path planner. 
The main difference with CBS is that the constraints in \AOCCBS are posed in continuous time, and the low-level path planner uses SIPP~\cite{SIPP} to work in continuous time.
This search forms the lower bound on the optimal cost; just like CBS, it searches optimistically from the bottom up, starting without any conflict-avoiding constraints and progressively adding them until a conflict-free solution is produced. Since constraints are only ever added during expansions, the lowest-cost candidate solution over all CT leaf nodes provides a lower bound on the optimal cost.
Unlike its predecessors, \AOCCBS runs a portfolio of policies that select and expand CT nodes in parallel, thereby using multiple CPU processors.
Each policy also applies a configurable set of repair functions that search for a potentially suboptimal incumbent solution. 
The best-found incumbent is made available even if the optimal has yet to be found, while also forming an upper bound on the optimal cost.

%% file: sections/classical_assumptions.tex
This section discusses the implications of the classical MAPF assumptions and qualitatively motivates why they are believed to worsen solution quality.
The discussion is set in the context of agents representing physical robots in a shared space, with conflicts representing collisions.
The assumptions discussed here are that time is discrete and conflicts are defined only by the agents' locations on the graph and not by their spatial positions or shapes.
From these assumptions come interconnected and limiting implications that concern both the validity (conflict-freeness) and the quality of a solution when applied to the real system.

\textbf{Synchronised timesteps induce artificial delays.}
Consider a graph with variable-length edges.
To maintain synchronicity among agents (as discrete time demands), 
agent~1 traversing a short edge must be delayed to match agent~2 traversing a longer edge.
Whether this delay on agent~1 comes via waiting somewhere or by reducing its speed, the cost increases nonetheless.
How such a delay is implemented affects the real agents' spatial positions and therefore requires care to preserve solution validity.
As clarified in the next points,
using only unit-length edges does not completely avoid these issues;
artificial delays will still exist
and solution validity depends on more than just edge lengths.


\textbf{Graph-based conflicts induce artificial delays.}
In the common square grid map layout with unit-length edges and circular agents~\cite{SturtevantBenchmarks},
consider two agents with radius~$r$, where agent~1 moves to agent~2's vertex while at the same time agent~2 moves away.
If agent~2 moves along an edge perpendicular to agent~1's direction, then agent~1 could start moving $1-2\sqrt{2}r$ earlier than agent~2 without colliding.
Taking this to the limit with agent~1 leaving at the same time as agent~2 gives the maximum radius under the classical formulation: 
$1-2\sqrt{2}r = 0 \Leftrightarrow r = \sqrt{2}/4$.
On the other hand, if agent~2 moves along the edge parallel to agent~1's direction, then agent~1 can begin its traversal $1-2r$ earlier than agent~2.
Thus, in this parallel case, even with the classical formulation's limit $r=\sqrt{2}/4$, agent~1 could move $1-2r \approx 0.29$ time units earlier.
The smaller $r$ is, the earlier agent~1 can traverse its edge relative to agent~2.
The artificial delays induced by classical MAPF not considering agent shape therefore grow as the agents get smaller.

\textbf{Graph-based conflicts limit the graph resolution.}
With discrete time, unit-length edges, and circular agents, 
agent radius must satisfy $r \leq \sin(\alpha/2) / 2$ units, 
where $\alpha$ is the smallest angle between any two edges in the graph.
In the square grid case, $\alpha=90^\circ \Rightarrow r\leq\sqrt{2}/4$.
If the agents are any larger than this limit then the classical solution is not guaranteed to be valid.
This practically means that the resolution at which the graph represents the physical environment is upper-bounded by the size of the agents.
A coarser graph permits only coarser movements, leaving less freedom in how the agents are coordinated.
In some cases, the size of existing agents and the required graph resolution may outright prohibit the use of classical MAPF. 

\textbf{Graph-based conflicts inhibit spatially overlapping edges}.
Since conflicts are defined entirely by agent locations on the graph, 
no two edges can spatially overlap without risking invalidity. 
This prohibits, e.g., the inclusion of both diagonals of a cell: 
since two agents traversing each respective diagonal at the same time will be
at the centre at exactly the same time, a collision occurs regardless of their shape.
Classical MAPF does not consider this a conflict since the agents are on different edges.
One workaround could be to only include one of the diagonals, thereby including at least some of these shorter moves. 
This together with time being discrete, however, 
leads to the same issue of synchronicity-induced delays as discussed in the first point of this section.
It would also reduce the maximum allowed agent radius for validity, due to the smaller angle between edges.

\begin{figure}
  \centering
  \begin{subfigure}[b]{0.32\linewidth}
    \centering
    \includegraphics[width=\linewidth]{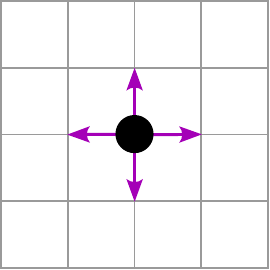}
    \caption{4-connected}
    \label{fig:Connectedness:4}
  \end{subfigure}
  \hfill
  \begin{subfigure}[b]{0.32\linewidth}
    \centering
    \includegraphics[width=\linewidth]{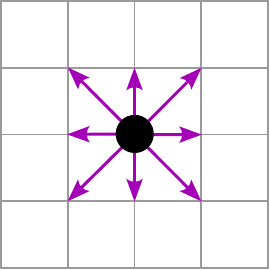}
    \caption{8-connected}
    \label{fig:Connectedness:8}
  \end{subfigure}
  \hfill
  \begin{subfigure}[b]{0.32\linewidth}
    \centering
    \includegraphics[width=\linewidth]{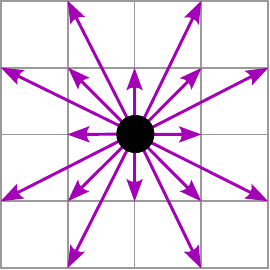}
    \caption{16-connected}
    \label{fig:Connectedness:16}
  \end{subfigure}
  \caption{Three levels of map connectedness. 4-connectedness includes only cardinal edges, 8-connectedness adds diagonals, and 16-connectedness adds knight's moves.}
  \label{fig:Connectedness}
  \Description{A representation of a grid map with three different levels of connectedness, known as $2^k$-neighbourhoods, determining which direct vertex-to-vertex movements are available. Each additional level of connectedness doubles the number of edges outgoing from a vertex.}
\end{figure}
These limitations of classical MAPF surely reduce its achievable solution quality.
After all, a diagonal between two opposing corners of a cell is ${\sim}29\%$ shorter than the Manhattan distance in 2D, and ${\sim}42\%$ shorter in 3D.
Fig.~\ref{fig:Connectedness} shows three levels of connectedness (known as $2^k$-neighbourhoods~\cite{rivera2017grid}), going from only cardinal edges (4), to adding diagonals (8), to adding knight's moves (16).
The improvements to cost with increasing connectedness are however diminishing, and is bounded by the cost of each agent travelling the Euclidean distance from their respective start to goal vertices.
For instance, going from 4 to 8-connectedness shortens a knight's move by $19.5\%$, 
but going from 8 to 16-connectedness shortens the same move by only an additional $7.4\%$.
Nonetheless, the potential in cost reduction from even only adding diagonals is significant, which classical MAPF forfeits.

Classical MAPF requires unit-length edges, but nothing limits this to square grids.
A triangular lattice (see Fig.~\ref{fig:Lattices:triangular}) is, so to speak, 6-connected since there are six different outgoing directions from a vertex,
possibly allowing agents to move more directly toward their goal vertices than with 4-connectedness (Fig.~\ref{fig:Lattices:square}).
This representation must still respect validity limits on agent size relative to the edge length, which is even more constrained for triangles than squares due to the smaller angle between edges.
We elaborate further on this in Section~\ref{sec:Discussion}, though we only investigate square lattices in this study.

\subsection{Can Map Resolution Recover the Loss?}
\label{sec:ClassicalAssumptions:Refinement}

Considering that classical MAPF solution-validity upper-bounds the graph resolution based on agent size, a natural question that may arise for systems with small agents is:
\emph{If agents are small, can classical MAPF achieve the same solution quality as \MAPFR by refining the map with smaller edges?}
Our main experiments in Section~\ref{sec:Results} keep the map static while reducing the size of the agents, which does not test this.
In practice, if the agents are small then the map can be made finer until the radius of the agents becomes $\sqrt{2}/4$ of an edge length.
Using a fine map can only improve the cost, since a solution on a coarse grid is also a solution on a fine grid (assuming the coarse edges are split evenly into fine edges).
We argue that the artificial delays induced by agent synchronicity and not considering agent shapes can partly be mitigated by using finer maps, up to the point when the graph cannot be made finer without risking 
solution validity.
However, no amount of refinement can achieve the gains from using non-cardinal edges in the map.
Manhattan distance is, after all, the same regardless of how fine the grid is.

%% file: sections/study_design.tex
The purpose of this study is to 
compare the SOC of optimal classical MAPF and \MAPFR solutions over 
(i) discrete and continuous time,
(ii) various agent sizes, and 
(iii) various levels of graph connectedness.
Since the classical formulation only works in discrete time, does not exploit agent shapes, and is not able to handle non-unit-length edges, 
we only search for the optimal classical MAPF solution on 4-connected maps, implicitly using agents with radius $r=\sqrt{2}/4$, 
which is then compared with the various \MAPFR optima along (i)--(iii).
Comparisons are done on different maps, with the aim of gaining a crude understanding of how graph topology affects the 
comparison.
To this end, we use CBS for obtaining an optimal cost \DTopt of a classical MAPF solution, 
and \AOCCBS for obtaining a lower bound \CTLB and an upper bound \CTUB on the optimal cost of a \MAPFR solution.
We report the bounds $\RedLB$ and $\RedUB$ on the true cost change $\Red \in \left[ \RedLB, \RedUB \right]$
that an optimal \MAPFR solution returns relative to an optimal classical MAPF solution, where
\begin{equation*}
    \RedLB = 100\cdot\left(\frac{\CTLB}{\DTopt}-1 \right)\quad\text{and}\quad \RedUB = 100\cdot \min\left( 0, \frac{\CTUB}{\DTopt}-1 \right).
\end{equation*}
Both bounds are multiplied by $100$ to be read in percent. \RedUB is clipped to $0$ since any classical solution is also a \MAPFR solution; 
the optimal \MAPFR solution can only improve on the classical solution.
Therefore, we already know that the optimal \MAPFR cost is less than or equal to $\DTopt$ and therefore that $\CTUB=\DTopt$ is a valid upper bound, hence $\RedUB\leq0\%$. 
However, our repository does provide the unclipped results to allow practitioners to compare these two specific algorithms (CBS and \AOCCBS) in terms of solution quality \emph{and} runtime performance, which is otherwise outside the scope of this work.

We report the mean \RedLB and \RedUB, minimum \RedLB, and maximum \RedUB.
For a given map and agent count, 
these statistics are taken over the set of scenarios (see Section~\ref{sec:MapsAndScenarios}) for which both \RedLB and \RedUB have been found at every combination of agent radii and connectedness.
This is to ensure a fair comparison across these values.
In practice, 
\AOCCBS always returns \CTLB, 
and \RedLB is known to be at most $0\%$,
so this set of commonly solved scenarios depends entirely on CBS returning an optimal solution.
We therefore also report CBS's success rate at solving scenarios at each agent count, indicating the amount of data support for the corresponding bounds.
The number of agents is increased until the success rate falls below $40\%$ (reporting only on those above) or until the spread of \Red becomes too large to provide any useful information.

\subsection{Evaluation Dimensions}
Isolating time from agent size is not entirely possible since continuous time is required to exploit agent shapes, and agent shapes are required for continuous time else the agents can be moved arbitrarily close to each other.
For this reason, the closest we get in this work to a pure comparison between discrete and continuous time is on 4-connected graphs with agent radius $r=\sqrt{2}/4$.
These tests show to what extent \MAPFR can avoid delays induced by agent movement being synchronised and their shapes not being considered.

To evaluate the impact of agent size on solution quality, we sweep over agent radii $r \in \left\{\sqrt{2}/16, \sqrt{2}/8, \sqrt{2}/4 \right\}$.
The sweep is limited to at most $\sqrt{2}/4$ since any larger would compromise the validity of the classical MAPF solutions, while lower radii would not.
As discussed in the previous section, the classical formulation enforcing synchronised actions and using shape-agnostic conflict definitions incurs artificial delays. 
With smaller agents, these delays grow relatively larger. 
We therefore aim to understand how large the impact of agent size is.

The impact of graph connectedness on solution quality is evaluated by using the three levels of connectedness shown in Fig.~\ref{fig:Connectedness}: 4-connected (cardinal movements), 8-connected (+ diagonals), and 16-connected (+ knight's moves).
Non-cardinal edges are only added if an agent with $r=\sqrt{2}/4$ can traverse it without intersecting a non-traversable cell;
these same edges are used for all agent radii.
The addition of more edges yields a richer set of move actions that allow both for shorter paths and a greater coverage of the space for more options when avoiding collisions. 

The above dimensions are cross-evaluated to understand their combined effect on solution quality.

\subsection{Maps and Scenarios}
\label{sec:MapsAndScenarios}

\begin{figure}
    \centering
    \noindent\hfill%
    \begin{subfigure}[b]{0.3\linewidth}
        \centering
        \includegraphics[width=\linewidth]{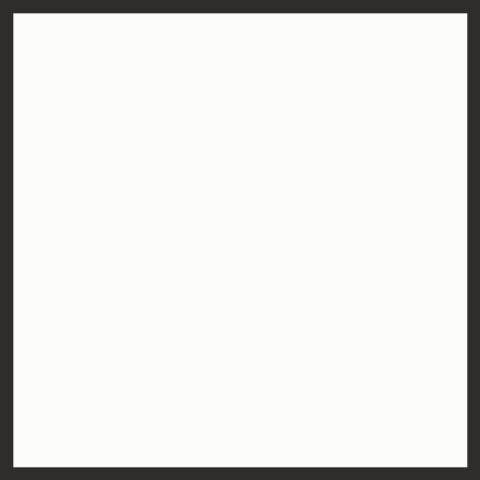}
        \caption{\mapEmpty}
        \label{fig:maps:empty32}
    \end{subfigure}%
    \hfill%
    \begin{subfigure}[b]{0.3\linewidth}
        \centering
        \includegraphics[width=\linewidth]{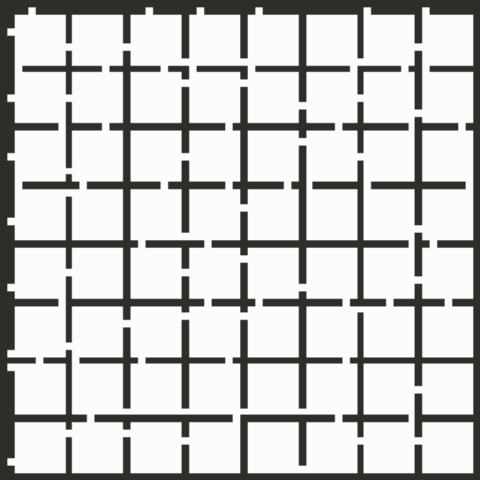}
        \caption{\mapRoom}
        \label{fig:maps:rooms}
    \end{subfigure}%
    \hfill\null

    \vspace{1em}

    \noindent\hfill%
    \begin{subfigure}[b]{0.61\linewidth}
        \centering
        \includegraphics[width=\linewidth]{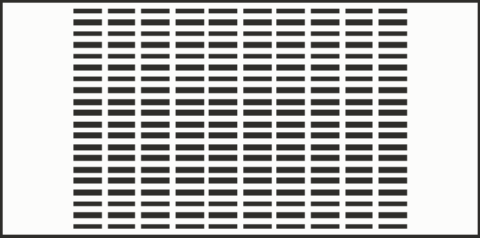}
        \caption{\mapWarehouse}
        \label{fig:maps:warehouse}
    \end{subfigure}%
    \hfill%
    \begin{subfigure}[b]{0.3\linewidth}
        \centering
        \includegraphics[width=\linewidth]{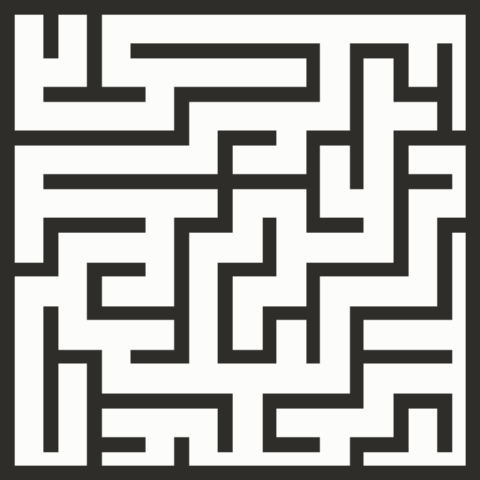}
        \caption{\mapMaze}
        \label{fig:maps:maze}
    \end{subfigure}%
    \hfill\null
    \caption{The four MovingAI~\cite{SturtevantBenchmarks} maps on which our main experiments compare the two formulations. White cells are traversable, black cells are not.}
    \label{fig:maps}
    \Description{A visualisation of the four MovingAI maps \mapEmpty, \mapRoom, \mapWarehouse, and \mapMaze, which are used in the main experiments. These maps consist of cells, some traversable and some not. The composition of these cells define the type and difficulty of planning agents in it. For instance, \mapMaze contains many narrow corridors which require more coordination between the agents than, say, \mapEmpty which is an empty and open space.}
\end{figure}

Our main experiments are done on four maps from the MovingAI benchmark suite~\cite{SturtevantBenchmarks}: \mapEmpty, \mapRoom, \mapWarehouse, and \mapMaze, shown in Fig.~\ref{fig:maps}. 
These maps provide a crude progression from open spaces to narrow corridors and choke-points.
Large open spaces allow for more non-cardinal edges and a larger number of equivalent-cost paths for agents to take. 
They also lead to more localised conflict avoidance since agents in general only need short delays to avoid other agents.
Narrow corridors and tight spaces, on the other hand, do not allow for as many non-cardinal edges, and they also incur higher delays due to larger mutually exclusive areas. 
For instance, an agent must wait for another agent to clear an entire corridor before it can itself move through it.

We use the 25 \emph{random} scenarios provided by~\cite{SturtevantBenchmarks} for each map: 
a scenario defines a list of start-goal pairs; 
a problem with $n$ agents is constructed by assigning to the agents the first $n$ start-goal pairs in the scenario.
An exception is made for \mapEmpty which instead uses scaled \mapEmptySmall scenarios, to keep results on these two maps comparable for the experiments described in the next section.

\subsection{Leveraging Map Resolution}
\label{sec:StudyDesign:Refinement}

Based on Section~\ref{sec:ClassicalAssumptions:Refinement}, we aim to test how much of \MAPFR's solution quality the classical MAPF formulation can recover by using a finer map resolution. 
To do so, we compare solution quality 
on \mapEmpty (a $32\times32$ grid) with \mapEmptySmall (a $16\times16$ grid);
\mapEmpty has twice the resolution.
Practically, we use agent radius $\sqrt{2}/4$ on \mapEmpty and $\sqrt{2}/8$ on \mapEmptySmall,
and use \mapEmptySmall's random scenarios on both maps but with start-goal positions scaled by $2$ and solution costs by $1/2$ when applied to \mapEmpty.
The problems on both maps are therefore equivalent, except that the effective graph resolution is doubled on \mapEmpty, 
allowing for agents to take steps of half the length compared to on \mapEmptySmall. 
We therefore say that \mapEmpty is the finer map, while \mapEmptySmall is the coarser.

\subsection{Settings and Hardware}

All experiments are run on AMD EPYC 9755 Turin 2.7 GHz 128-core processors, with each run using at most 4-cores and 16~GB of DDR5 memory. 
Both CBS and \AOCCBS are given a 300~second runtime limit.
We use the default CBS settings as provided in the implementation by~\cite{LI2021103574}.
For \AOCCBS, we use a four-policy (4~CPU core) gap-cadence portfolio, 
which prioritises finding solutions by applying repair functions while the optimality gap upper bound is above $5\%$, and thereafter gradually reduces repair function use to instead focus on raising \CTLB.
Our repository provides detailed settings for each algorithm.

%% file: sections/results.tex
\begin{figure*}
    \centering
    \includegraphics[width=1\textwidth]{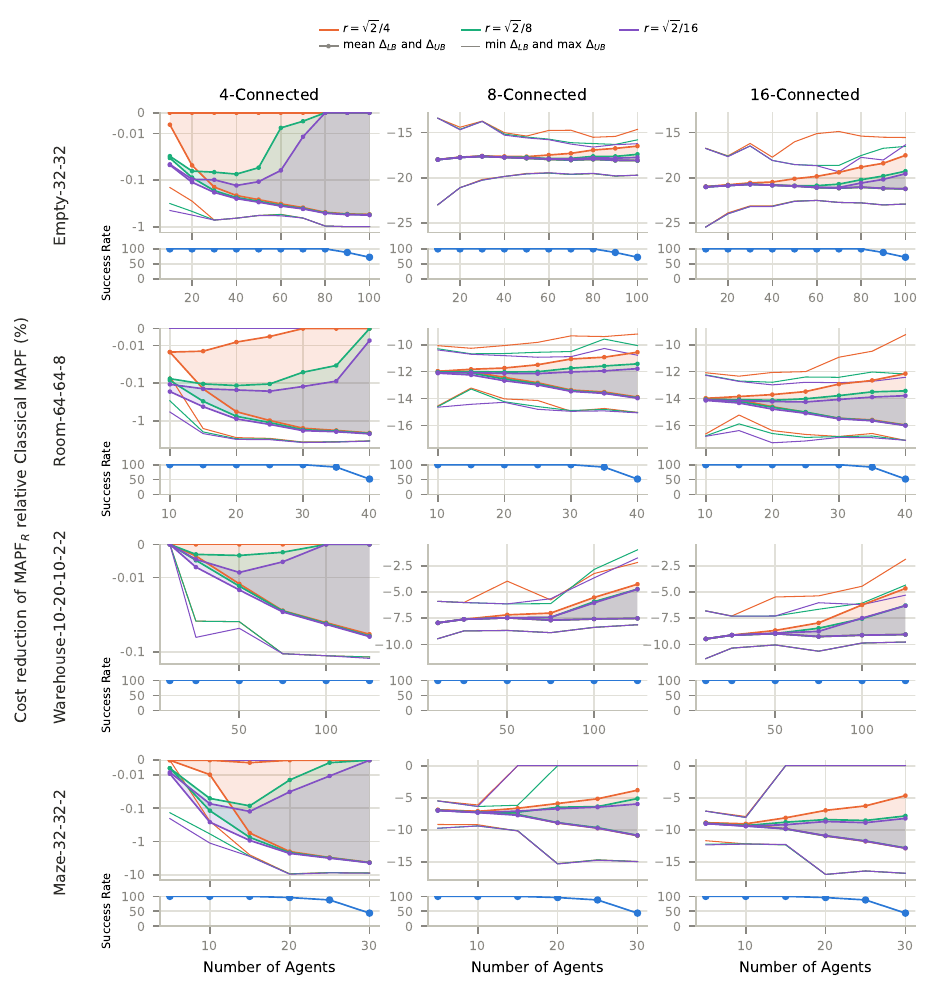}
    \caption{The solution cost reduction of \MAPFR relative to classical MAPF, on four maps (rows) with different connectedness (columns) and agent radius (colours). Each plot shows mean bounds on the cost change, $\Red\in[\RedLB, \RedUB]$, together with the minimum \RedLB and maximum \RedUB, in percentage. The y-axes for the 4-connected graphs are symmetric-logarithmic while the 8- and 16-connected are linear. Each map has its own agent count on the x-axes. CBS's success rate is shown below each corresponding plot.}    
    \label{fig:MainResults}
    \Description{The results from the main experiments, showing the change in cost from using \MAPFR instead of the classical formulation, with upper and lower bounds on the mean change, and the minimum and maximum change included. This figure shows how the difference in solution quality between classical MAPF and \MAPFR changes depending on the context.}
\end{figure*}

Fig.~\ref{fig:MainResults} shows the cost reduction of \MAPFR relative to classical MAPF on each of the maps, connectedness levels, and agent radii.
For a given graph corresponding to a specific map and connectedness, each colour corresponds to a specific agent radius, for which there are four lines:
the two thicker lines show the mean $\RedLB$ and $\RedUB$ over scenarios, bounding the shaded area within which the true value of $\Red$ is;
the two thinner lines above and below (overlapping in some places) show the maximum $\RedUB$ and minimum $\RedLB$ over scenarios;
and the dots show which agent counts were tested.
CBS's success rate over scenarios is shown under each subplot. 

To summarise the main findings before taking a closer look:
on 4-connected maps, 
\RedLB extends down to at most only a few percent, 
meaning that the cost reduction of \MAPFR over classical MAPF is relatively small at these tested agent counts. 
That is, the direct improvement from time being continuous and agent shapes being considered is limited,
despite smaller agents allowing for lower-cost solutions (most clearly seen in Fig.~\ref{fig:MainResults} at lower agent counts).
The results instead show that the value of \MAPFR comes from admitting higher levels of connectedness, which improves the cost change by at least an order of magnitude.  
The advantages from higher connectedness and smaller agents appear to be additive, each contributing to tighter $[\RedLB, \RedUB]$ bounds at lower costs.  
Across the maps, we see the improvements from higher connectedness decreasing as the open spaces become less, 
which is expected since narrower maps allow for fewer non-cardinal edge connections.

\textbf{(i) Discrete versus continuous time:}
The 4-connected maps with $r=\sqrt{2}/4$ provide our cleanest comparison between discrete and continuous time, without the effects of smaller agents or higher map connectedness.
Keep in mind, however, that agent shapes are still considered, meaning that this is not a pure discrete/continuous time comparison. 
The results show \RedLB reaching down to at most only a few percent, meaning that the \MAPFR formulation does relatively little to improve the solution quality at these agent counts.
The spread of \Red remains relatively high throughout the agent count range, which means that \AOCCBS for these specific cases is unable to narrow the bounds by much within the given runtime.

\textbf{(ii) Smaller agents:}
Smaller agent radii generally improve both \RedLB and \RedUB across all maps and connectedness levels.
The cases at low agent counts where \RedUB for smaller agents is below \RedLB for larger agents empirically confirms that smaller agents allow for higher-quality solutions.
\RedUB corresponds to actual \MAPFR solutions found by \AOCCBS;
the fact that smaller agent radii decreases \RedUB means that smaller agents may make it easier for \AOCCBS to find solutions; 
the same may be true for other \MAPFR algorithms.
This makes intuitive sense since collisions become less likely when agent sweep-volumes are smaller.
However, \AOCCBS finding lower-cost solutions is confounded with the fact that smaller agents permit lower-cost solutions, 
and so we refrain from drawing any more conclusions regarding the impact of agent size on the ease of finding solutions.

With smaller agents, the spread of \Red is narrower at the lower end of the agent range,
revealing a trend of decreasing \Red with more agents. 
This suggests that \MAPFR pays off with more agents.
However, since \AOCCBS is unable to maintain sufficiently narrow bounds as the number of agents increases,
we are not able to draw any further conclusions regarding how far this trend of decreasing \Red goes.
We leave these questions for future investigations using more scalable \MAPFR algorithms than \AOCCBS.

\textbf{(iii) Map connectedness:}
Increasing map connectedness has by far the largest impact on solution quality out of all our tested dimensions. 
Going from 4-connectedness to 16-connectedness on \mapEmpty improves $\Red$ from fractions of a percent to over $17\%$ on average. 
This is due to the expanded set of move actions shortening paths and offering more alternatives for conflict avoidance.
The results show that the largest increase can already be achieved with 8-connectedness, while 16-connectedness provides moderate further improvements.
The reduction in cost from higher connectedness tapers off over the maps \mapEmpty$\rightarrow$\mapRoom$\rightarrow$\mapWarehouse$\rightarrow$\mapMaze, when qualitatively going from most to least open spaces.
However, even on \mapMaze with long and narrow corridors, the average improvement is still significant.

\subsection{Impact of Higher Map Resolution}
\begin{figure}
  \centering
  \includegraphics[width=1\linewidth]{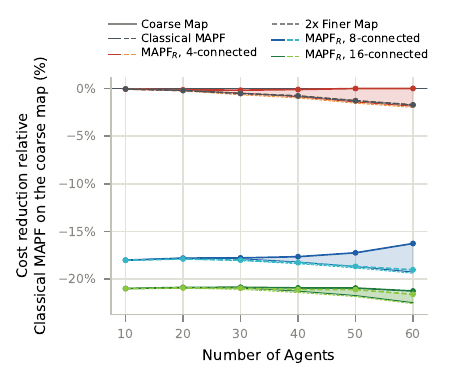}
  \caption{
  The solution cost reduction (in percent) from using classical MAPF on the fine map, or \MAPFR with various connectedness levels on both the coarse and fine map, compared to the baseline of classical MAPF on the coarse map. For each setting, the dotted line shows \RedUB (with dots showing specifically which agent counts were tested), a corresponding thinner line showing \RedLB, and the shared area showing where the true reduction $\Red\in[\RedLB, \RedUB]$ must be.}
  \label{fig:Refinement}
  \Description{The results from refining the map to twice the resolution, using both the classical and \MAPFR formulations. The is figure therefore shows how staying within the classical MAPF formulation and double the resolution of the map gives relatively little in return, while using the \MAPFR formulation significantly improves the solution quality when increasing the connectedness. Increasing both connectedness and the map resolution together result in even higher-quality solutions.}
\end{figure}

Fig.~\ref{fig:Refinement} visualises the cost improvement from doubling the map resolution.
The \MAPFR 4-connected solutions are clipped to the solution quality of the classical MAPF solutions on the same map, since they are also a solution to the \MAPFR problem.
By doubling the map resolution, with classical MAPF or \MAPFR, a cost reduction of less than $3\%$ is achieved for these agent counts.
It is not possible to increase the resolution to achieve further improvements without compromising the validity of the classical MAPF solutions; 
this is therefore the extent of the improvements that are possible through a higher map resolution in this context.
These gains are marginal in comparison to increasing the map connectedness, 
which reduces the cost by more than $16\%$ with 8-connectedness and $21\%$ with 16-connectedness.
These results mirror those in Fig.~\ref{fig:MainResults}: 
here we show that using a twice as fine map improves costs slightly more than what continuous-time and shape-awareness do on their own, but still relatively little compared to using higher map connectedness.
Interestingly, \RedLB is almost identical across resolutions, however, \RedUB on the fine map (most pronounced with 4 and 8-connectedness) is significantly closer to \RedLB.
This means that using a finer map does not help \AOCCBS raise \CTLB that much, but it seems to help \AOCCBS find lower-cost solutions and thereby lower \CTUB.

%% file: sections/discussion.tex
The results show that continuous time and shape-awareness do improve solution quality on their own,
but only by a few percent at the agent counts we tested.
Higher connectedness improves it by at least an order of magnitude more.
This is not to say that continuous-time or shape-awareness are not needed,
in actuality they are necessary for enabling non-unit-length and spatially overlapping edges.
A cardinal edge and a diagonal respectively have lengths $1$ and $\sqrt{2}$, which no common timestep divides,
so a formulation with synchronised timesteps cannot represent both faithfully.
Besides, since classical MAPF defines conflicts by graph location, 
the two diagonals of a cell cannot both be included without risking the solution's validity.
Adding diagonals therefore requires relaxing both the discrete-time and graph-based conflict assumptions.
This is also why, to our knowledge, no classical MAPF algorithm incorporates non-unit-length edges.
Therefore, the true cost of the classical MAPF assumptions is mainly paid through ruling out a richer set of move actions.

The experiments over different map topologies show that classical MAPF forfeits a significant part of the achievable solution quality on maps with open spaces, where many non-cardinal edges would otherwise be possible.
This loss is more modest on maps with narrow corridors and tight spaces.
The classical formulation therefore remains a reasonable simplification on these more constrained maps,
especially given how scalable and robust its algorithms have become after decades of research effort.
If \MAPFR algorithms achieve the level of scalability that real-world applications require, then the additional solution quality that they unlock becomes practically attainable.
Our results also show that much of the solution quality is attained already at 8-connectedness; for this reason research effort may be best spent on these lower connectedness levels.

\begin{figure}
  \centering
  \begin{subfigure}[b]{0.30\linewidth}
    \centering
    \includegraphics[width=\linewidth]{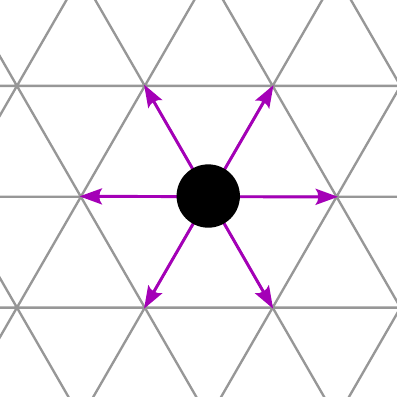}
    \caption{Triangle}
    \label{fig:Lattices:triangular}
  \end{subfigure}
  \hfill
  \begin{subfigure}[b]{0.30\linewidth}
    \centering
    \includegraphics[width=\linewidth]{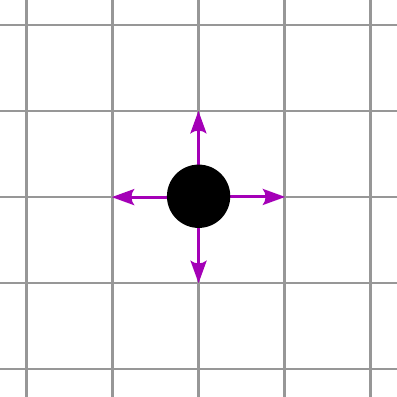}
    \caption{Square}
    \label{fig:Lattices:square}
  \end{subfigure}
  \hfill
  \begin{subfigure}[b]{0.30\linewidth}
    \centering
    \includegraphics[width=\linewidth]{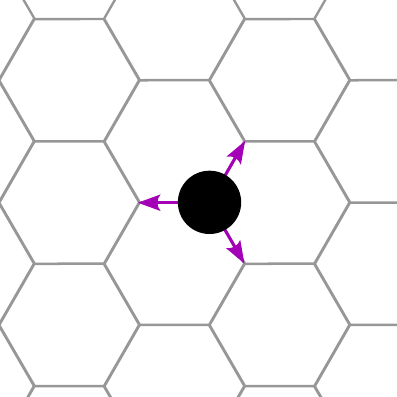}
    \caption{Hexagon}
    \label{fig:Lattices:hexagon}
  \end{subfigure}
  \caption{Three types of lattices, with the agent size fixed and the edge lengths set to the minimum that classical MAPF validity permits.}
  \label{fig:Lattices}
  \Description{A visualisation of different lattices (triangular, square, and hexagonal), all which the classical MAPF formulation can be applied to. The edge lengths are set to their minimum possible without compromising a solution's validity.}
\end{figure}

Furthermore, 
although MAPF research often uses \emph{square} lattice graphs, classical MAPF does not require that.
Higher-quality solutions may be obtained through other lattices, such as the triangular or hexagonal lattices in Fig.~\ref{fig:Lattices}.
We acknowledge the practical difficulty of this, 
since many real-world settings are built around right angles.
Nonetheless, with agent radius limited to  $r \leq \sin(\alpha/2) / 2$ where $\alpha$ is the smallest angle between adjacent edges in the graph,
this gives for triangular lattices $r\leq0.25$ ($\alpha=60^\circ$),
for square lattices $r\lesssim 0.35$ ($\alpha=90^\circ$), 
and for hexagonal lattices $r\lesssim0.43$ ($\alpha=120^\circ$).
Larger permitted agent radii mean that a finer map resolution can be used;
although the actual density of vertices is also decided by the lattice's shape.
The shape also determines the connectedness: six for triangles, four for squares, three for hexagons. 
This presents a trade-off, where more connected lattices force longer edges.
However, as our results show, 
connectedness has a far larger impact on solution quality than map resolution.
We leave investigations into the trade-off between different lattices to future work.

Another interesting direction, based on connectedness being the strongest improver of solution quality, is MAPF with Asynchronous Actions~\cite{CBSAA, ren2021loosely}.
There, classical MAPF's discrete-time assumption is relaxed to allow for non-unit-length edges, while graph-based conflicts are still kept.
The latter means that agent sizes and edge lengths are still bounded to each other for solution validity,
and overlapping edges are still prohibited.
However, since it is more similar to the classical MAPF formulation than what \MAPFR is, 
it is possible that many of the algorithmic developments extending classical MAPF scalability could be transferred to this formulation.
Although, for the time being we have yet to see it applied to agent counts of the same magnitude as what classical MAPF has been applied to~\cite{ren2021loosely, wu2026bcbs, CBSAA}.

\AOCCBS returns valid bounds on the optimal \MAPFR cost, so the direction and validity of our results do not depend on which specific algorithm was used.
What does depend on it is the width of these bounds and the agent count that can be reached while keeping these bounds narrow enough to offer insights.
This is clearest on the map \mapMaze with $r=\sqrt{2}/4$ and 4-connectedness, where at relatively small agent counts the bounds extend from $0\%$ down to $-10\%$, effectively offering very little insight.
We suspect the cause of this to be long corridors,
which are known to be particularly difficult for CBS-style algorithms.
The CBS implementation used here~\cite{LI2021103574} uses specific techniques to address this, however, corresponding techniques have to our knowledge not yet been developed for continuous-time and are therefore not implemented in \AOCCBS.
Narrower bands would tell us whether or not \Red keeps decreasing as the agent count grows, thereby answering the question regarding whether the cost reduction from continuous-time and shape-awareness alone eventually become significant or not.
Since \MAPFR generalises classical MAPF, which is NP-hard~\cite{NPhardness}, as far as we know there will always be an agent count beyond which the bounds widen;
a more scalable algorithm than \AOCCBS only moves that point further out.

This study uses only 2D grid maps.
A 3D diagonal is even shorter than a 2D diagonal when compared to the corresponding Manhattan distance.
Both classical MAPF and \MAPFR can also be applied to 3D; therefore, comparing these two formulations in 3D space would make for interesting future work.
Furthermore, 2D grid maps do not exercise the full representational power of \MAPFR.
\emph{Roadmaps}, which need not follow any regular structure, can be adaptively tuned to have a higher resolution in parts of the environment where it is needed, such as at junctions and high-traffic areas in the environment. 
This freedom is not present in classical MAPF since map resolution is limited to agent size.
Our results should therefore be read as a conservative estimate of what the classical formulation forfeits.
Finally, classical MAPF plans for synchronous systems and generally needs post-processing before non-holonomic robots can execute them~\cite{honig2016multi, ConcreteMAPP}.
\MAPFR supports kinodynamically feasible trajectories directly~\cite{AOCCBS},
so the cost gap we measure is likely a lower bound on the gap in execution.

%% file: sections/conclusion.tex
Classical MAPF has been the dominant MAPF formulation for many years,
with the scalability of modern algorithms enabling the planning of thousands of agents.
However, classical MAPF fundamentally assumes discrete time and graph-based conflicts, thereby forfeiting solution quality.
This work investigated the extent of this forfeit by comparing optimal classical MAPF solutions to those of Continuous-time MAPF (\MAPFR), a formulation that relaxes several of the classical MAPF assumptions.
A comparison was done over discrete and continuous time, various agent sizes, and graph connectedness, resolution, and topology.
At our tested agent counts, continuous-time and shape-awareness alone account for marginal improvements;
however, they enable an expanded set of move actions that lead to significant improvements in solution quality.
Average improvements on narrow and constrained maps are at least $5\%$, on open maps at least $17\%$, and in some cases more than $20\%$.
We also find that naively refining the map, 
which cannot be done unboundedly in classical MAPF without compromising the solution's validity, only slightly raises solution quality. 
Classical MAPF therefore cannot recover what its assumptions forgo through more compute. 
We do however pinpoint contexts where the forfeit is only moderate: 
maps with narrow corridors and tight spaces, where agents are relatively large compared to the graph edges, and where only cardinal movements are possible. 
This work therefore quantifies what classical MAPF loses through its assumptions, which had previously only been partially observed or hinted at.